\documentclass{elsart}
\newfont{\cal}{cmsy10}
\usepackage{epsfig}
\usepackage[latin1]{inputenc}

\newcommand{\eps}{\varepsilon}
\newcommand{\bfr}{\begin{flushright}}
\newcommand{\efr}{\end{flushright}}

\begin{document}

\begin{frontmatter}
\title{Top quark loop corrections to the decay 
   $H^+ \rightarrow h^0 W^+$ \\
   in the Two Higgs Doublet Model }
\author{R. Santos${}^1$, A. Barroso}
\address{Dept. de F\'\i sica, Faculdade de Ci\^encias, Universidade de Lisboa \\
         Campo Grande, C1, 1700 Lisboa\thanksref{em1}}
\author{L. Brücher\thanksref{inida}}
\address{Institut f\"ur Physik, Johannes Gutenberg-Universit\"at\\
         Staudingerweg 7, D-55099 Mainz\thanksref{em2}}

\thanks[jnict]{Partially supported by JNICT contract BD/2077/92-RM}
\thanks[em1]{e-mail: fsantos@skull.cc.fc.ul.pt}
\thanks[inida]{Partially supported by INIDA}
\thanks[em2]{e-mail: bruecher@dipmza.physik.uni-mainz.de}

\begin{abstract}
We calculate the decay width for the process $H^+ \rightarrow h^0 W^+$
up to order $g^4$ in the framework of the Two Higgs Doublet Model. We 
argue that for some reasonable choice of the free parameters the contribution
from the one-loop graphs can be as large as 80\%.
\end{abstract} 

\end{frontmatter}

\section{Introduction}

Despite the enormous success of the $SU(2)\otimes U(1)$ Electroweak theory,
the fundamental mechanism responsible for the gauge boson masses remains 
untested. This by itself justifies the study of extensions of the minimal 
model. Among these various extensions the most important one is the 
two-Higgs doublet model (2HDM). In fact, even without considering 
supersymmetry, the existence of more than one generation of scalar fields  
is a possibility that ought to be explored. (See ref.\cite{hunt} for a general 
review).

The existence of charged scalar particles $H^+ , H^-$ is the cleanest 
signature for the 2HDM. so it is important to study the production and decays 
of these particles. The production of $H^+ H^-$ is kinematically suppressed in 
lepton colliders. On the contrary, in hadron colliders one could produce a 
substantial number of charged Higgs via the reaction 
$ g \bar{b} \rightarrow H^+ \bar{t} $ \cite{gun}. After the production, the
dominant decay channel is $H^+ \rightarrow t \bar{b}$ which, unfortunately,
due to the large QCD background, makes the detection very difficult. For this 
reason, the alternative channel $H^+ \rightarrow h^0 W^+$, where $h^0$ is the
lightest of the neutral Higgs bosons, could be very important. The calculation 
of the top quark loops to the decay width of the process 
$H^+ \rightarrow W^+ h^0$ is our aim.

Previously, some of us \cite{rui1,rui2} have discussed the vacuum
stability and the renormalization of the most general 2HDM with CP
conservation. There are two kinds of potentials with only CP invariant
minima and both dependent on seven real parameters. In here, we work
with the following potential, denoted by $V(I)$ in ref.\cite{gun},
\begin{equation}
  \label{pot}
  V = -\mu_1^2 x_1 -\mu_2^2 x_2 + \lambda_1 x_1^2 + \lambda_2 x_2^2
        + \lambda_3 x_3^2  + \lambda_4 x_4^2  + \lambda_5 x_1 x_2
\end{equation}
where
\begin{equation}
  x_1=\phi_1^+\phi_1,\qquad  x_2=\phi_2^+\phi_2,\qquad  
  x_3=\mbox{Re}(\phi_1^+\phi_2),\qquad  x_4=\mbox{Im}(\phi_1^+\phi_2)
\end{equation}
and $\phi_i$ are two complex scalar doublets with hypercharge Y=1.

To renormalize the model using on-shell prescription the seven real
parameters of $V$ are replaced by the square of the vacuum expectation
value $v^2=v_1^2+v_2^2$, the masses of the Higgs particles, $H^+$, 
$H^-$, $H^0$, $h^0$ and $A^0$, the ratio $\frac{v_2}{v_1} \equiv \tan
\beta$ and the angle $\alpha$, which rotates the mass eigenstates
$H^0$ and $h^0$ to the $SU(2)$ eigenstates.

\section{The decay width}

Let $q$ denote the 4-momentum of $H^+$, $q'$ the 4-momentum of $h^0$
and $p=q-q'$ the 4-momentum of $W^+$. Thus, at tree level the decay
amplitude $H^+ \rightarrow h^0 W^+$ is 
\begin{equation}
  T=\eps^*_{\mu} \Gamma_0^{\mu}
\end{equation}
with
\begin{equation}
  \Gamma_0^{\mu} = \frac{g}{2} \cos (\beta - \alpha)(q+q')^{\mu}
\end{equation}
This, in turn leads to the following expression to the decay width
\begin{equation}
  \Gamma = \frac{g^2 \cos^2(\beta - \alpha)}{64 \, \pi M_W^2
  M_{H^+}^3} \,\left[ \left( M_{H^+}^2 -  M_{h^0}^2 - M_{W}^2
  \right)^2 - 4  M_{h^0}^2 M_{W}^2 \right]^{\frac{3}{2}}
\end{equation}

\begin{figure}[htbp]
  \begin{center}
    \leavevmode
     \epsfig{file=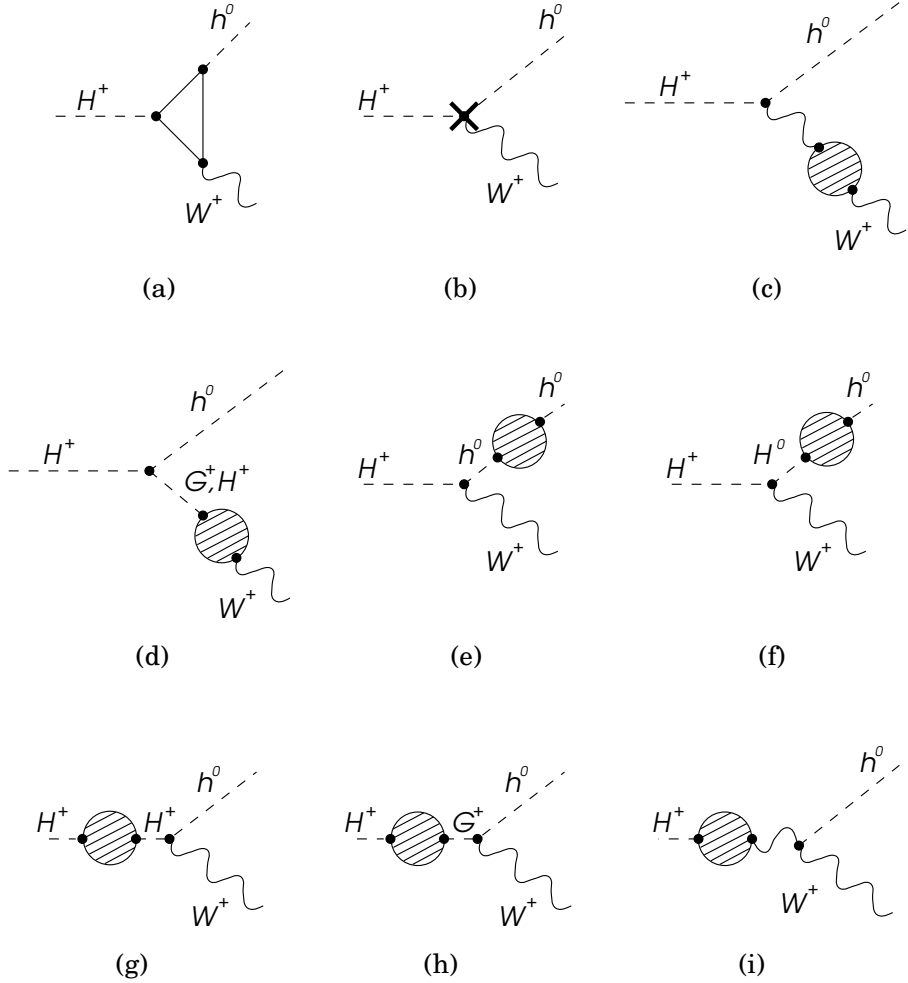,width=12cm}        
    \caption{Feynman graphs at one-loop level}
    \label{fig:feyn}
  \end{center}
\end{figure}

At one-loop order, the renormalized $H^+ H^0 W^+$ vertex
$\Gamma^{\mu}_{1ren}$ is represented in fig.~\ref{fig:feyn}. 
Diagrams a) and b) are
the unrenormalized proper vertex and its counterterm, respectively,
and the remaining diagrams, where the crossed circles represent the
renormalized two-point Greens functions, are corrections to the
external legs.

Using the on-shell renormalization prescription all these diagrams
vanish, assuming, as we do, that the particles are on-shell. Without
giving the details that can be found elsewhere \cite{rui2,aoki}, let
us simply make a few comments. Clearly the on-shell condition implies
that diagram c) vanishes. Similarly, diagram d) is also zero, because,
for an on-shell $W^+$, the mixed self-energy is proportional to
$p^{\mu}$ and $\eps \cdot p = 0$.

Under renormalization, particles with the same quantum numbers get
mixed. So the relation between the bare fields $h^0$ and $H^0$, for
instance, and the renormalized ones is a matrix, i.e.,
\begin{equation}
\left[ 
  \begin{array}{l}
    H^0 \\ h^0
  \end{array} \right]_{0} \, = \,
\left[ 
  \begin{array}{ll}
    Z^{\frac{1}{2}}_{H^0 H^0} &  Z^{\frac{1}{2}}_{H^0 h^0} \\ 
    Z^{\frac{1}{2}}_{h^0 H^0} &  Z^{\frac{1}{2}}_{h^0 h^0} 
  \end{array} \right]
\left[ 
  \begin{array}{l}
    H^0 \\ h^0
  \end{array} \right], \quad 
M_{i0}^2=M_i^2+\delta M_i^2, \quad  i=H^0, h^0.
\end{equation}
Then the wave function renormalization constants are fixed by the
following conditions, imposed on the renormalized self-energies
\renewcommand{\theequation}{\arabic{equation}\mbox{a}}
\begin{equation}
 \left. \Sigma_R^{H^0 H^0} (p^2) \right|_{p^2=M_{H^0}^2} \, = \, 0
\end{equation}
and
\addtocounter{equation}{-1}
\renewcommand{\theequation}{\arabic{equation}\mbox{b}}
\begin{equation}
  \left. \Sigma_R^{H^0 h^0} (p^2) \right|_{p^2=M_{h^0}^2} \, = \, 0
\end{equation}
These conditions guarantee that diagrams e) and f) vanish, Similar
conditions can be imposed on the renormalized self-energy of the
charged Higgs and on the mixing self-energy $H^+ G^+$ between the
charged Higgs and the Goldstone boson $G^+$, i.e.
\renewcommand{\theequation}{\arabic{equation}\mbox{a}}
\begin{equation}
  \left. \Sigma_R^{H^+ H^+} (p^2) \right|_{p^2=M_{H^+}^2} \, = \, 0
\end{equation}
and
\addtocounter{equation}{-1}
\renewcommand{\theequation}{\arabic{equation}\mbox{b}}
\begin{equation}
  \left. \Sigma_R^{H^+ G^+} (p^2) \right|_{p^2=0} \, = \, 0 .
\end{equation}
This, in turn, guarantees that diagrams g) and h) are also zero. Now,
the counterterm for the $H^+ W^+$ is given without any further
constraint by
\renewcommand{\theequation}{\arabic{equation}}
\begin{equation}
  -\frac{i}{2} p^{\mu} M_W \delta Z_{G^+H^+}
\end{equation}
where $\delta Z_{G^+H^+}$ is the off diagonal term for the wave
function renormalization matrix of the charged scalars. Then, the
vanishing of diagram i) provides a consistency check of the
cal\-cu\-la\-tion.

\begin{figure}[htbp]
  \begin{center}
    \leavevmode
     \epsfig{file=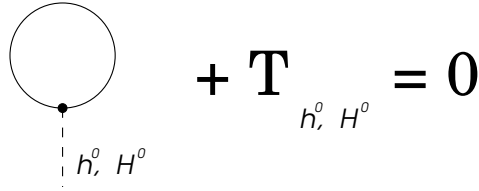,height=2.5cm}    
    \caption{tadpole renormalization condition.}
    \label{fig:tads}
  \end{center}
\end{figure}

For the sake of completeness, we write the counterterm Lagrangian
$\mbox{\cal{L}}_{CT}$ for the CP-even charged Higgs sector, namely
\begin{eqnarray}
 \mbox{\large\cal L}_{C.T.} & = & -\delta m_{H^0}^2  +  (q^2 - m_{H^0}^2) \delta
 Z_{H^0 H^0} - \delta_{H^0 H^0} \nonumber \\
 &  & -\delta m_{h^0}^2  +  (q^2 - m_{h^0}^2) \delta
 Z_{h^0 h^0} - \delta_{h^0 h^0} \nonumber \\
 &  &  +  \frac{1}{2}(q^2 - m_{H^0}^2) \delta Z_{H^0 h^0} 
  +   \frac{1}{2}(q^2 - m_{h^0}^2) \delta Z_{h^0 H^0} - \delta_{H^0 h^0} \\
 &  & -\delta m_{H^+}^2  +  (q^2 - m_{H^+}^2) \delta
 Z_{H^+ H^+} - \delta_{H^+ H^+} \nonumber \\
 &  &  +  q^2 \delta Z_{G^+ G^+} - \delta_{G^+ G^+} \nonumber \\
 &  &  +   \frac{1}{2}(q^2 - m_{H^+}^2) \delta Z_{H^+ G^+} 
  +  \frac{1}{2} q^2 \delta Z_{G^+ H^+} - \delta_{H^+ G^+} \nonumber
\end{eqnarray}
where 
\begin{eqnarray}
  \delta_{H^0 h^0} & = & \frac{\sin 2\alpha}{v\sin 2\beta}\left(
  T_H \sin(\alpha-\beta) + T_h\cos(\alpha-\beta)\right) \nonumber \\ 
  \delta_{H^0 H^0} & = & \frac{2}{v\sin 2\beta}\left( T_H
  (\cos^3\alpha\sin\beta+\sin^3\alpha\cos\beta) \right. \nonumber \\
  & & \qquad \qquad \qquad \left. 
    + T_h\sin\alpha\cos\alpha\sin(\alpha-\beta)\right) \nonumber \\ 
  \delta_{h^0 h^0} & = & \frac{2}{v\sin 2\beta}\left( T_H
  \sin\alpha\cos\alpha\cos(\alpha-\beta) \right. \nonumber \\
  & & \qquad \qquad \qquad \left.  +T_h
  (\cos^3\alpha\cos\beta-\sin^3\alpha\sin\beta)\right) \nonumber \\
  \delta_{H^+ G^+} & = & \frac{1}{v}\left(
  T_H \sin(\alpha-\beta) + T_h\cos(\alpha-\beta)\right) \\ 
  \delta_{G^+ G^+} & = & \frac{1}{v}\left(
  T_H \cos(\alpha-\beta) - T_h\sin(\alpha-\beta)\right) \nonumber \\ 
  \delta_{H^+ H^+} & = & \frac{2}{v\sin 2\beta}\left(
  (\cos^3\beta \sin\alpha + \sin^3\beta\cos\alpha) T_H  \right. \nonumber \\
  & & \qquad \qquad \qquad \left. 
  +(\cos^3\beta\cos\alpha - \sin^3\alpha\sin\beta) T_h \right)  \nonumber
\end{eqnarray}
$T_H$ and $T_h$ are the tadpole counterterms, fixed by the
renormalization condition on the 1-particle Green function, as shown
in fig.\ref{fig:tads}.

The calculation of diagram a) is standard. There are several particles
that can be included in the loop. However, in here we consider quark
loops, assumed to be dominant due to the large top quark mass. Notice
that this subset has to be finite by itself.

Diagram b) of fig. 1 has two main contributions. The first one comes
from the parameter variation on the tree level coupling $H^+ H^0 W^+$
and it is 
\begin{eqnarray}
  \label{vari}
\lefteqn{  i\frac{g}{2} \,\cos(\beta -\alpha) \,(q+q')^{\mu}\,\left[\; \frac{\delta g}{g} 
  \; + \; \frac{1}{2}\delta Z_{H^+ H^+} \; + \; \frac{1}{2}\delta Z_{h^0 h^0} \right.} \nonumber \\
  & & \qquad \qquad \qquad \qquad \qquad \quad \left .  
  + \; \frac{1}{2}\delta Z_{W W} \; - \; \tan(\beta-\alpha) \delta(\beta-\alpha)\; \right]
\end{eqnarray}
The second contribution is due to the existence of other tree level
couplings $H^+ H^0 W^+$ and $G^+ h^0 W^+$, that, due to wave function
mixing, induced the vertex that we want, namely:
\begin{equation}
  \label{vert1}
  i\frac{g}{2} \,\sin(\beta-\alpha)\,(q+q')^{\mu}\,\left[
  \frac{1}{2}\delta Z_{H^+ G^+} -\frac{1}{2}\delta Z_{h^0 h^0} \right]
\end{equation}

Besides $\delta(\beta-\alpha)$, all parameter in eq.(\ref{vari}) and
(\ref{vert1}) are already fixed. In fact, like in the minimal standard
model, $\delta g$ id fixed by the photon electron vertex. The
interesting point to notice is that one requires a
$\delta(\beta-\alpha)$ to obtain a renormalized finite result. In
principle, in this model $\beta$ and $\alpha$ are two physical
parameters that could be fixed independently. However, for
illustrative purpose we have  decided to renormalize the angle
$(\beta-\alpha)$ using the similar process $H^+ \rightarrow H^0 W^+$
which at tree level is proportional to $\sin(\beta-\alpha)$. Then,
$\delta(\beta-\alpha)$ is fixed imposing that the one-loop $H^+ H^0
W^+$ vertex vanishes for $q^2=M_{H^+}^2$, $q'^2=M_{H^0}^2$ and
$p^2=M_W^2$, i.e.
\begin{equation}
  \label{ren1}
  \left. \Gamma^{\mu}_{1ren}\left(H^+ H^0 W^+\right)
  \right|_{(M_{H^+}^2, M_{H^0}^2, M_W^2 )} \, = \, 0
\end{equation}
This implies that, in this model, without further measurements, the
decay $H^+ \rightarrow H^0 W^+$ fixes the parameter $(\beta-\alpha)$
and $H^+ \rightarrow h^0 W^+$ checks the consistency of the theory at
one-loop level. Clearly this implies a kinematical bound 
\begin{displaymath}
  M_{H^+}^2 > M_W^2+M_{H^0}^2
\end{displaymath}

It is interesting to point out that our calculation is similar to the
study of the radiative decay $H^+ \rightarrow \gamma W^+$ \cite{ray}.
In that case, the main one-loop corrections are also due to top quark
loops, but there is a major difference. As a consequence of the
electromagnetic $U(1)$ invariance, there is no tree level
contribution. Then, the proper vertex counterterm ( the equivalent of
fig 1b) ) cancels with the counterterms of the external legs diagrams
\cite{rui3}. This means that the calculation can be done simply by
summing all unrenormalized reducible and irreducible diagrams. This
sum is finite and electromagnetic gauge invariant. On the contrary,
the one-loop calculation for the decay width $H^+ \rightarrow h^0 W^+$
requires a detailed renormalization program for the 2HDM.

\section{Results and discussion}

For our numerical calculation we used a computer Maple\cite{Map}
program\cite{loop}. All quark masses except $m_t = 180$ GeV and $m_b =
4.1$ GeV have been neglected. The values of the remaining parameters
were taken from Particle Data Group\cite{data}, the CKM matrix element
$V_{tb}$ was set equal to one and the angles $\beta$ and $\alpha$ were
varied in the range $0<\beta<\frac{\pi}{2}$ and
$-\frac{\pi}{2}<\alpha<\frac{\pi}{2}$. According to tree unitarity
analysis \cite{tranal} the Higgs boson masses are bounded from 
above as $M_{H^+} < 870$ GeV,
$M_{H^0} < 710$ GeV and $M_{h^0} < 500$ GeV. In our numerical examples
we respect these bounds but no further constraints are imposed.

\begin{figure}[htbp]
  \begin{center}
    \leavevmode
    \epsfig{file=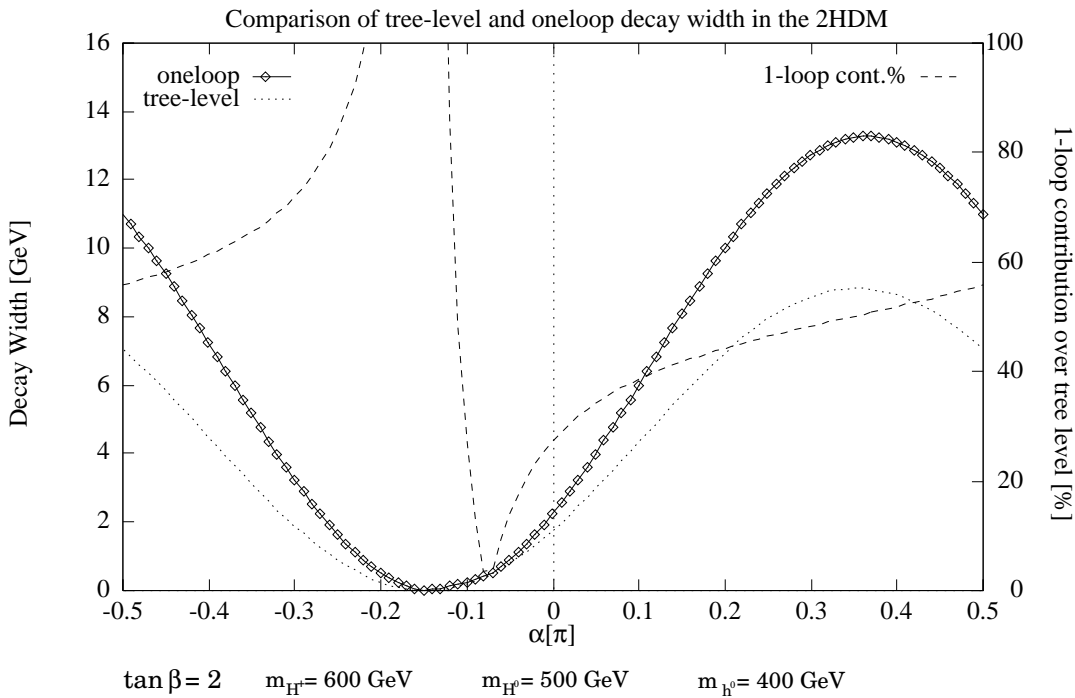,height=9.2cm}
    \caption{Decay width as a function of $\alpha$.
      One-loop means the contribution from the tree level plus the 
      contribution from the one-loop graphs.}
    \label{plot1}
  \end{center}
\end{figure}

In fig. \ref{plot1} we show the decay width for $M_{H^+} = 600$ GeV, 
$M_{H^0} = 500$ GeV, $M_{h^0} = 400$ GeV and $\tan\beta = 2$ as a
function of $\alpha$. The dotted curve gives the tree level result,
while the full curve includes also top quark loops. Depending on the
value of $\alpha$, the one-loop result varies between 20\% and 60\% of
the tree level result. Obviously this enhancement depends on the
values of the parameters. The dashed curve shows the relative
importance of the one-loop contribution in percentage. This curve
grows to infinity when $\alpha = \beta - \frac{\pi}{2}$ which
corresponds to a zero tree level result.  

\begin{figure}[htbp]
  \begin{center}
    \leavevmode
    \epsfig{file=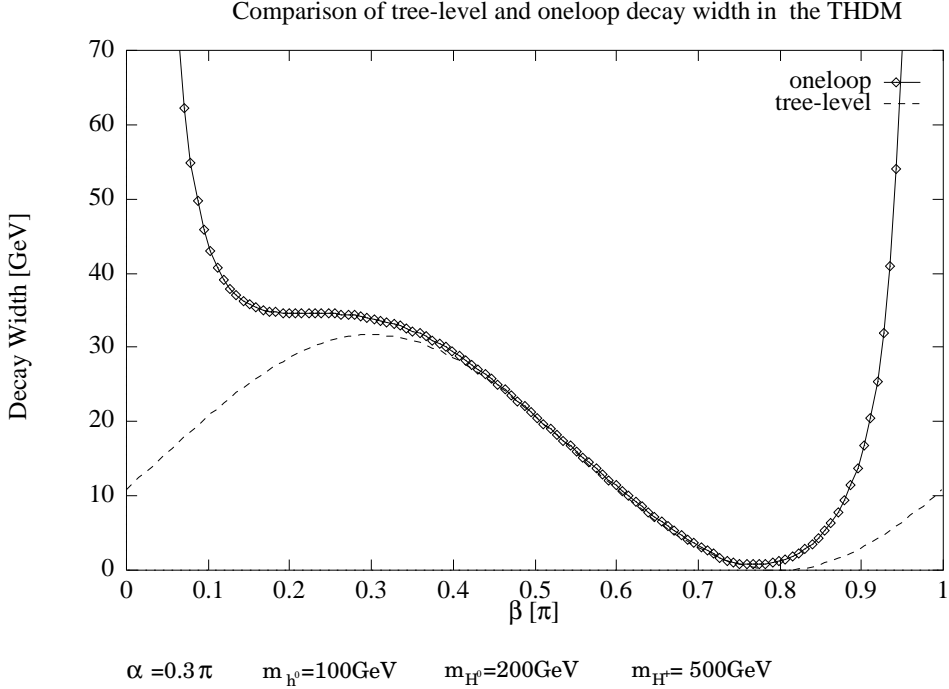,height=9.2cm}
    \caption{Decay width as a function of $\beta$.}
    \label{plot2}
  \end{center}
\end{figure}

The range that we have indicated is representative of reasonable values
for the angles. However, very large enhancement can be obtained for
small values of $\beta$. This is shown in fig. \ref{plot2}, where we
plot the decay width as a function of $\beta$ for $M_{H^+} = 500$ GeV,
$M_{H^0} = 200$ GeV, $M_{h^0} = 100$ GeV and $\alpha = 0.3\pi$. Notice
that a small $\beta$ implies a very large coupling between the Higgs
and the top quark. Obviously, at some point, perturbation theory breaks
down.

\begin{figure}[htbp]
  \begin{center}
    \leavevmode
    \epsfig{file=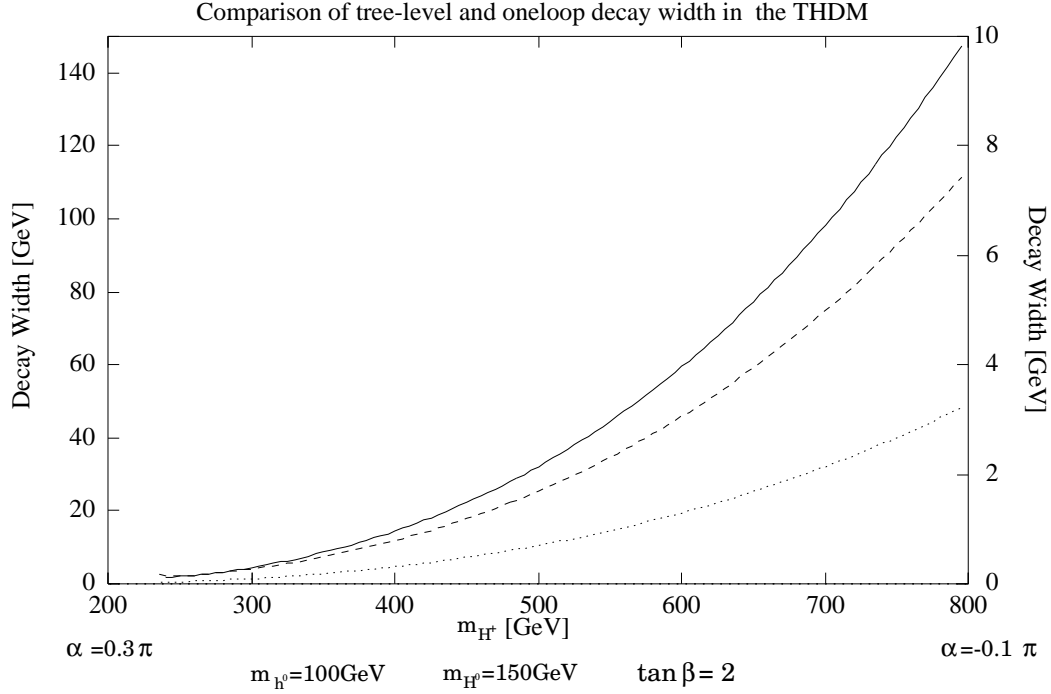,height=9.2cm}
    \caption{Decay width as a function of $M_{H^+}$ for
      $\alpha=-0.1\pi$(dashed and dotted curve) and $\alpha=0.3\pi$(solid).}
    \label{plot3}
  \end{center}
\end{figure}

Finally in fig. \ref{plot3} we show the variation of the decay width
with the mass of the charged Higgs boson, for $M_{H^0} = 150$ GeV, 
$M_{h^0} = 100$ GeV, $\tan\beta = 2$ and two values for $\alpha$,
$\alpha = 0.3\pi$ and $\alpha = -0.1\pi$. As we would expect, the
decay width of $H^+ \rightarrow h^0 W^+$ grows with $M_{H^+}$ and the
relative importance of the quark loop corrections also grows with
$M_{H^+}$. In this figure, the right hand scale corresponds to the
dashed and dotted curves which are the tree level(dotted) and tree
plus one-loop (dashed)
for $\alpha = -0.1\pi$. The solid curve has the scale on the
left side and represents the tree level plus one-loop result for
$\alpha = 0.3\pi$. On the same scale, the tree level curve is almost
coincident with the solid curve and for this reason it is not
shown. We have also studied the dependence of the result in
$M_{h^0}$. Leaving aside the phase space dependence, which is most
important at threshold, the dependence is mild, and there is no point
showing it. The same happens with the dependence on $M_{H^0}$.

The comparison of these two cases illustrates the following qualitative
argument: the width is larger when $\alpha \approx \beta$, but in this
case the loop corrections are smaller (about 1\% for any value of
$M_{H^+}$), on contrary, when $\beta$ and $\alpha$ are different the
overall result is smaller but the quark loop corrections grow in
relative importance, reaching in some cases 80\% of the tree level
contribution.


\begin{thebibliography}{99}

\bibitem {hunt} J. Gunion, H. Haber, G. Kane, S. Dawson \\
                The Higgs Hunter's Guide, Addison Wesley (1990)
\bibitem {gun}  J. Gunion, H. Haber, F. Paige, Wu-ki Tung 
                and S.S.D Willenbrock \\
                Nucl. Phys. B294 (1987) 621 
\bibitem {rui1} J. Velhinho, R. Santos and A. Barroso \\
                Phys. Lett. B322 (1994) 213
\bibitem {rui2} A. Barroso and R. Santos, in preparation
\bibitem {aoki} K. Aoki, Z. Hioki, R. Kawabe, M. Konuma, T. Muta \\
                Suppl. Prog. Theor. Phys. 73 (1982) 1
\bibitem {ray}  S. Raychaudhuri and A. Raychaudhuri, \\
                Phys. Lett B297 (1992) 159
\bibitem {rui3} J. Soares and A. Barroso \\
                Phys.Rev. D39 (1989) 1973
\bibitem {Map} B. W. Char, K. O. Geddes, G. H. Gonnet, B. L. Leong,\\
               M. B. Monagan, S. M. Watt: Maple V, Springer (1991)
\bibitem {loop} L. Br\"ucher, J. Franzkowski, D. Kreimer. \\
                Computer Physics Communication 85 (1995) 153-165
\bibitem {data} L. Montanet et al. \\
                Phys. Rev. D 50, 1173 (1994) and 1995 off-year partial
                update for the 1996 edition \\
                ( available via WWW http://pdg.lbl.gov/ )
\bibitem {tranal} S. Kanemura, T. Kubota and E. Takasugi, \\
                Phys. Lett. B 313, 155 (1993);\\
                J. Maalampi, J. Sirkka, I. Vilja, \\
                Phys. Lett. B 265, 371 (1991).

\end{thebibliography}
\end{document}